\documentclass[useAMS,usenatbib]{mn2e} 
 \usepackage{graphicx}

 \title[Detection of periodic variations]
 {Detection of periodic variations in vertical velocities of Galactic masers}
 \author[V. V. Bobylev and A. T. Bajkova]{V. V. Bobylev$^{1,2}$
 \thanks{E-mail: vbobylev@gao.spb.ru} and A. T. Bajkova$^{1}$\\
 $^{1}$Central (Pulkovo) Astronomical Observatory of RAS, 65/1 Pulkovskoye Chaussee, St.~Petersburg, 196140, Russia\\
 $^{2}$ Sobolev Astronomical Institute, St. Petersburg State University, Bibliotechnaya pl. 2, St.~Petersburg, 198504, Russia}

\begin{document}
 \date{Accepted 2014 MONTH XX.} 
 \pagerange{\pageref{firstpage}--\pageref{lastpage}} \pubyear{2014}
 \maketitle
 \label{firstpage}

\begin{abstract}
We have collected literature data on Galactic masers with
trigonometric parallaxes measured by means of VLBI. We have
obtained series of residual tangential, $\Delta V_{circ}$, and
radial, $\Delta V_R$, velocities for 107 masers. Based on these
series, we have re-determined parameters of the Galactic spiral
density wave using the method of spectral (periodogram) analysis.
The tangential and radial perturbation amplitudes are
   $f_\theta=6.0\pm2.6$~km s$^{-1}$ and
        $f_R=7.2\pm2.2$~km s$^{-1}$, respectively; the perturbation
 wavelengths are
   $\lambda_\theta=3.2\pm0.5$~kpc and
   $\lambda_R=     3.0\pm0.6$~kpc for a four-armed spiral model, $m=4$.
The phase of the Sun
   $\chi_\odot$ in the spiral density wave is
   $ -79^\circ\pm14^\circ$ and
   $-199^\circ\pm16^\circ$ from the
residual tangential and radial velocities, respectively. The most
interesting result of this work is detecting a wave in vertical
spatial velocities ($W$) versus distance $R$ from the Galactic
rotation axis. From the spectral analysis, we have found the
following characteristics of this wave: the perturbation
wavelength is
 $\lambda_W=3.4\pm0.7$~kpc and the amplitude
       $f_W=4.3\pm1.2$~km s$^{-1}$.
\end{abstract}


 \section{INTRODUCTION}
Like many other galaxies, our Galaxy is characterized by a spiral
structure. There exist numerous models explaining the origin and
maintenance of the spiral pattern. Proposed models include the
swing amplification of noise
 \citep{Goldreich1965,Julian1966}, groove modes
 \citep{Sellwood1989,Sellwood1991,Sellwood2012} or other modes
 \citep{Lin1964}, possibly recurrent
 \citep{Sellwood-Garl2014}.

Most models of the Galaxy's spiral structure consider only
perturbations of radial and tangential velocities in the Galactic
plane. It is interesting to note that large-scale, coherent
vertical motions have been found in the disc of the Milky Way in
SEGUE \citep{Widrow2012}, RAVE \citep{Williams2013} and LAMOST
\citep{Carlin2013} data.

Non-zero vertical velocities of objects are usually explained with
impact of external objects, like dwarf galaxies (companions) or
clouds of dark matter, crossing the Galactic disk. However, it is
possible to explain them without involvement of any external
forces. In the framework of the theory of spiral density waves,
small fluctuations can occur in the direction perpendicular to the
Galactic plane. For example, \cite{Fridman2007} pointed out the
possibility of such fluctuations. Numerical simulations made
recently by various authors (for example, \cite{Faure2014} and
\cite{Debattista2014}) showed that the spiral density wave in the
Galaxy might lead to the emergence of vertical fluctuations with
an average amplitude of 10--20~km~s$^{-1}$.

Galactic masers with known trigonometric parallaxes measured by
means of VLBI \citep{Reid09,Honma2012,Reid14} give unprecedented
opportunity to study fine kinematical effects. Relative
measurement errors of parallaxes provided by VLBI are, on average,
less than 10\%. Currently, the number of such objects is already
119. In this Letter, we re-determine parameters of the Galactic
spiral density wave from radial and tangential velocities of these
119 masers and investigate their vertical velocities for
periodicities which can be associated with the influence of the
Galactic spiral density wave.

\section{METHOD}\label{method}
From observations, we have heliocentric line-of-sight velocity
$V_r$ in $\mathrm{km\,s}^{-1}$;  proper-motion velocity components
$V_l=4.74r\mu_l \cos b$ and $V_b=4.74r\mu_b$, in the $l$ and $b$
directions respectively (the coefficient $4.74$ is the number of
kilometers in an astronomical unit divided by the number of
seconds in a tropical year); heliocentric distance $r$ in kpc for
a star. The proper motion components $\mu_l \cos b$ and $\mu_b$
are measured in $\mathrm{mas\, yr}^{-1}$. We adopt
$R_0=8.0\pm0.4$~kpc for the galactocentric distance of the Sun.

The components $U,V,W$ of stellar spatial velocities  are
determined from the observed line-of-sight and tangential
velocities in the following way:
 \begin{equation}
 \begin{array}{lll}
 U=V_r\cos l\cos b-V_l\sin l-V_b\cos l\sin b,\\
 V=V_r\sin l\cos b+V_l\cos l-V_b\sin l\sin b,\\
 W=V_r\sin b                +V_b\cos b.
 \label{UVW}
 \end{array}
 \end{equation}
Velocity $U$ is directed towards the Galactic center; $V$, along
the Galactic rotation; and $W$, towards the Northern Galactic
pole.

Two projections of these velocities: $V_R$, directed radially from
the Galactic center towards an object, and $V_{circ}$, orthogonal
to $V_R$ and directed towards Galactic rotation, are found from
the following relations:
 \begin{equation}
 \begin{array}{lll}
  V_{circ}= U\sin \theta+(V_0+V)\cos \theta, \\
       V_R=-U\cos \theta+(V_0+V)\sin \theta,
 \label{VRVT}
 \end{array}
 \end{equation}
where $V_0=|R_0\Omega_0|$, and the position angle $\theta$ is
determined as $\tan\theta=y/(R_0-x)$, where $x,y$ are Galactic
Cartesian coordinates of an object. The quantity $\Omega_0$ is the
Galactic angular rotational velocity at distance $R_0$; parameters
$\Omega^1_0, \ldots, \Omega^n_0$ are derivatives of the angular
velocity from the first to the $n$th order, respectively.

Velocities $V_R$ are practically independent and velocities $W$,
independent from the nature of the Galactic rotation curve.
However, to analyze periodicities in the tangential velocities, it
is necessary to find parameters of a smooth rotation curve and to
form residual velocities $\Delta V_{circ}$. As it is shown by
practice, to build a smooth rotation curve in a wide range of
distances $R$, it is sufficient to have values of two derivatives
of the Galactic rotation angular velocity: $\Omega^\prime_0$ and
$\Omega^{\prime\prime}_0$. All the three velocities: $V_R$,
$\Delta V_{circ}$ and $W$ must be corrected for the solar peculiar
velocity $U_\odot,V_\odot,W_\odot$.

To determine the angular rotational velocity $\Omega_0$ and its
two derivatives, $\Omega^\prime_0$ and $\Omega^{\prime\prime}_0$,
from observational data, we use the equations derived from
Bottlinger's formulas with the angular velocity of Galactic
rotation $\Omega$ expanded in a series to terms of the second
order of smallness in $r/R_0:$
\begin{equation}
 \begin{array}{lll}
 V_r=-U_\odot\cos b\cos l-V_\odot\cos b\sin l-W_\odot\sin b\\
 +R_0(R-R_0)\sin l\cos b\Omega^\prime_0\\
 +0.5R_0(R-R_0)^2\sin l\cos b\Omega^{\prime\prime}_0,
 \label{EQ-1}
 \end{array}
 \end{equation}
 \begin{equation}
 \begin{array}{lll}
 V_l= U_\odot\sin l-V_\odot\cos l-r\Omega_0\cos b\\
 +(R-R_0)(R_0\cos l-r\cos b)\Omega^\prime_0\\
 +0.5(R-R_0)^2(R_0\cos l-r\cos b)\Omega^{\prime\prime}_0,
 \label{EQ-2}
 \end{array}
 \end{equation}
 \begin{equation}
 \begin{array}{lll}
 V_b=U_\odot\cos l\sin b + V_\odot\sin l \sin b-W_\odot\cos b\\
    -R_0(R-R_0)\sin l\sin b\Omega^\prime_0\\
    -0.5R_0(R-R_0)^2\sin l\sin b\Omega^{\prime\prime}_0,
 \label{EQ-3}
 \end{array}
 \end{equation}
where the distance $R$ of an object from the Galactic rotation
axis is expressed as
  \begin{equation}
 R^2=r^2\cos^2 b-2R_0 r\cos b\cos l+R^2_0.
 \end{equation}
According to the linear theory of density waves \citep{Lin1964},
 \begin{equation}
 \begin{array}{lll}
       V_R =-f_R \cos \chi,\\
 \Delta V_{circ}= f_\theta \sin\chi,
 \label{DelVRot}
 \end{array}
 \end{equation}
where
 \begin{equation}
\chi=m[\cot(i)\ln(R/R_0)-\theta]+\chi_\odot
 \end{equation}
is the phase of the spiral wave ($m$ is the number of spiral arms;
$i$ is the pitch angle; $\chi_\odot$ is the radial phase of the
Sun in the spiral wave; $f_R$ and $f_\theta$ are amplitudes of
radial and tangential components of the perturbed velocities
which, for convenience, are always considered positive).

First, we use Bottlinger's equations (\ref{EQ-1})--(\ref{EQ-3})
only for construction of the Galactic rotation curve. Deviations
of tangential velocities from this curve give residual tangential
velocities $\Delta V_{circ}$. It is possible, however, to obtain a
solution for amplitudes of radial and tangential perturbations,
wavelength and phase of the Sun including (\ref{DelVRot}) into
Bottlinger's equations as shown by \cite{Mish1979}. Because the
vertical velocities $W$ of interest are not included to the model,
we next decided to use spectral analysis as unified approach to
study periodicities in all velocity sets: $V_R$, $\Delta V_{circ}$
and $W$ separately. Note that if radial and residual tangential
velocities correspond to the Lin\& Shu model (\ref{DelVRot}), then
separate and joint solutions should be the same. If results
differ, we can assume that there is some discrepancy between data
and the model (\ref{DelVRot}). Below we will compare separate and
joint solutions, obtained from the $V_R$ and $\Delta V_{circ}$
sets.

We use the effective method of spectral (periodogram) analysis
based on Fourier transform, modified for analysis of logarithmic
spirals and accounting for position angles of  the analyzed
objects \citep{Bajkova2012}.

The parameter $\lambda$, the distance (along the Galactocentric
radial direction) between adjacent spiral arm segments in the
Solar neighborhood (the wavelength of the spiral density wave), is
calculated from the relation:
\begin{equation}
 \frac{2\pi R_0}{\lambda} = m\cot(i).
 \label{a-04}
\end{equation}
Let there be a series of measured velocities $V_n(R_n)$ (these can
be both radial, $V_R$, and residual tangential, $\Delta
V_{\theta}$, velocities) at points with Galactocentric distances
$R_n$ and position angles $\theta_n$, $n=1,\dots,N$ where $N$ is
the number of objects. The objective of our spectral analysis is
to extract a periodicity from the data series in accordance with
the model describing a spiral density wave with parameters
 $f_R$ $(f_\theta$),
 $\lambda (i)$ and $\chi_\odot$.

Having taken into account the logarithmic character of the spiral
density wave and the position angles of the objects $\theta_n$,
our spectral (periodogram) analysis of the series of velocity
perturbations is reduced to calculating the square of the
amplitude (power spectrum) of the standard Fourier transform:
 \citep{Bajkova2012}:
\begin{equation}
 \bar{V}_{\lambda_k} = \frac{1} {N}\sum_{n=1}^{N} V^{'}_n(R^{'}_n)
 \exp\Bigl(-j\frac {2\pi R^{'}_n}{\lambda_k}\Bigr),
 \label{29}
\end{equation}
where $\bar{V}_{\lambda_k}$ is the $k$th harmonic of the Fourier
transform, with the wavelength $\lambda_k=D/k$; $D$ is the period
of the series being analyzed,
 \begin{equation}
 \begin{array}{lll}
 R^{'}_{n}=R_{\circ}\ln(R_n/R_{\circ}),\\
 V^{'}_n(R^{'}_n)=V_n(R^{'}_n)\times\exp(jm\theta_n).
 \label{21}
 \end{array}
\end{equation}
The algorithm of searching for periodicities modified to properly
determine not only the wavelength but also the amplitude of the
perturbations is described in detail in \cite{Bajkova2012}.

Obviously, the sought-for wavelength $\lambda$ corresponds to the
peak value of the power spectrum $S_{peak}$. The pitch angle of
the spiral density wave can be derived from Eq.~(\ref{a-04}). We
determine the perturbation amplitude and phase by fitting the
harmonic with the wavelength found to the observational data. The
following relation can also be used to estimate the perturbation
amplitude:
$$
f_R(f_\theta)=\sqrt{4\times S_{peak}}.
$$

\section{DATA}\label{Data}
We use coordinates and trigonometric parallaxes of masers measured
by VLBI with errors that are, on average, below 10\%. These masers
are related to very young objects (basically protostars of high
masses, but there are also those with low masses; a number of
massive supergiants are known as well) located in active
star-forming regions.

One of such observational campaigns is the Japanese project VERA
(VLBI Exploration of Radio Astrometry) for observations of water
(H$_2$O) Galactic masers at 22~GHz~\citep{Hirota07} and SiO masers
(such masers are very rare among young objects) at
43~GHz~\citep{Kim08}.

Water and methanol (CH$_3$OH) maser parallaxes are observed in the
USA (VLBA) at 22~GHz and 12~GHz~\citep{Reid09}. Methanol masers
are observed also in the framework of the European VLBI
network~\citep{Rygl10}. These two projects are combined in the
BeSSeL program~\citep{BesseL11}.

VLBI observations of radio stars in continuum at
8.4~GHz~\citep{Torres2007,Dzib11} are carried out with the same
goals. In the framework of this program, low-mass nearby stars
closely associated with the Gould belt are observed.

\cite{Reid14} present a summary of measurements of trigonometric
parallaxes, proper motions and radial velocities of 103 masers. To
this sample, six more sources in the Local arm were added from the
list in~\cite{Xu13}:
 EC~95,
 L~1448C,
 S1,
 DoAr21,
 SVC13/NGC1333,
 IRAS 16293$-$2422;
 then, we add two red supergiants,
 PZ~Cas \citep{Kusuno2013} and
 IRAS 22480$+$6002 \citep{Imai2012}, and
 finally,
 IRAS 22555$+$6213 \citep{Chibueze14},
 Cyg X-1~\citep{Reid11},
 IRAS 20143$+$3634 \citep{Burns2014}, and
five low-mass close radio stars in Taurus:
 Hubble~4 and
 HDE~283572 \citep{Torres2007},
 T~Tau~N   \citep{Loinard2007},
 HP~Tau/G2 \citep{Torres2009} and
 V773~Tau  \citep{Torres2012}.
 Thus, the whole sample contains 119 sources.

\section{RESULTS AND DISCUSSION}
We found the following components of the Solar peculiar velocity
and Galactic rotation parameters:
 \begin{equation}
 \label{OMEGA}
 \begin{array}{lll}
 (U_\odot,V_\odot,W_\odot)=(6.8,15.9,8.3)\pm(1.2,1.2,0.9){\rm~km~s^{-1}}\\
      \Omega_0 = ~29.53\pm0.49~\hbox{km s$^{-1}$ kpc$^{-1}$},\\
  \Omega^{'}_0 = -4.27\pm0.10~\hbox{km s$^{-1}$ kpc$^{-2}$},\\
 \Omega^{''}_0 = 0.874\pm0.050~\hbox{km s$^{-1}$ kpc$^{-3}$}.
 \end{array}
 \end{equation}
This solution has been obtained by least squares using 107 masers.
Following the recommendation of \cite{Reid14}, when determining
the Galaxy rotation parameters, masers located at distances
$R<4$~kpc were excluded because of the influence of the Central
Galactic bar at these distances. Then, several additional sources
were excluded on the base of the $3\sigma$ criterion:
 G09.62$+$0.19,
 G10.62$-$0.38,
 G12.02$-$0.03,
 G23.70$-$0.19,
 G25.70$+$0.04,
 G27.36$-$0.16,
 G78.12$+$3.63,
 G168.06$+$0.82 and
 G182.67$-$3.26.
In addition, we did not use the maser G28.86$+$0.06 that has a too
high velocity, $W>50$~km/s.

 \begin{figure}
 \includegraphics[width=1.0\columnwidth]{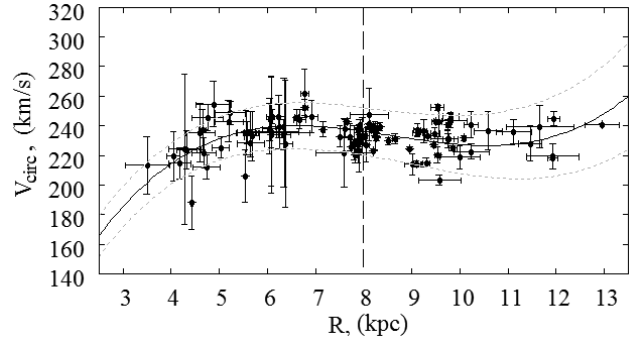}
 \caption{
The rotation curve of the Galaxy, built with the
parameters~(\ref{OMEGA}). The vertical line indicates the position
of the Sun. The borders of $1\sigma$ confidence intervals
correspond to the $\sigma_{R_\circ}=0.4$~kpc errors
 and errors of Galactic rotation parameters~(\ref{OMEGA}). }
  \label{f1}
 \end{figure}

\begin{figure}
 \includegraphics[width=1.0\columnwidth]{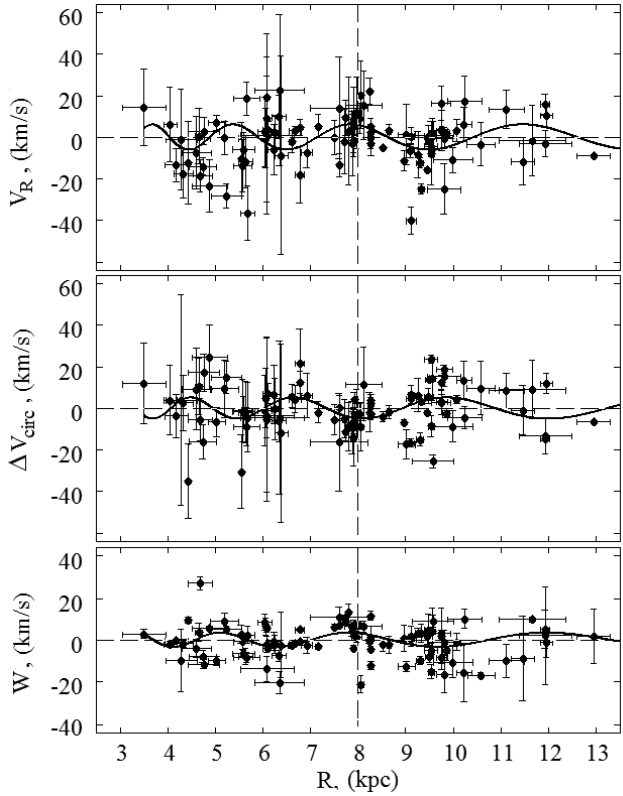}
 \caption{
The radial $V_R$, residual tangential $\Delta V_{circ}$ and
vertical velocities $W$ of masers versus distances $R$. The
vertical dashed line indicates the position of the Sun. }
  \label{f2}
 \end{figure}

\begin{figure}
 \includegraphics[width=1.0\columnwidth]{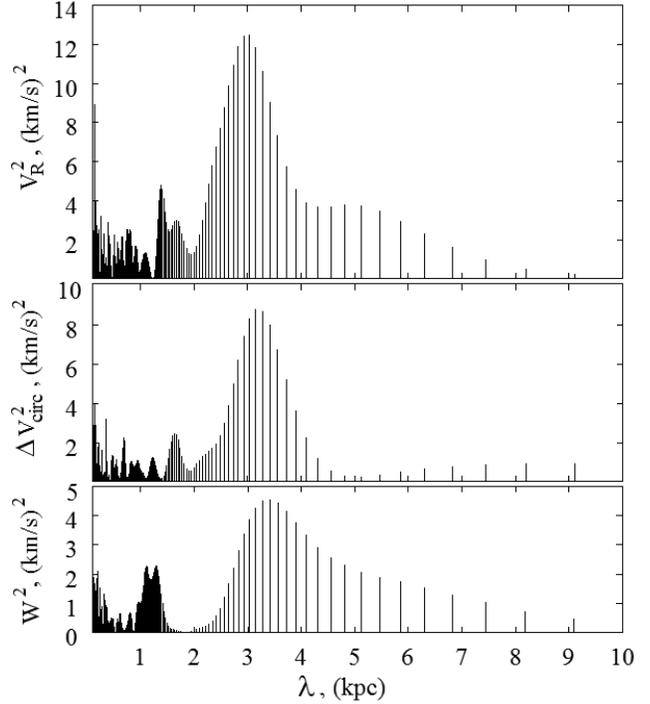}
 \caption{
Power spectra of the radial $V_R$, residual tangential $\Delta
V_{circ}$ and vertical $W$ velocities of masers. }
  \label{f3}
 \end{figure}

The Galaxy rotation curve, built with the
parameters~(\ref{OMEGA}), is presented in Fig.~\ref{f1}. Next, we
rejected 12~masers belonging to the Gould belt in order to avoid
the influence of its peculiar properties on the results of further
analysis. Velocities $V_R$, $\Delta V_{circ}$ and $W$ of 95 masers
versus $R$ are shown in Fig.~\ref{f2}. In this figure, a
periodicity in $V_R$ perturbations is clearly visible. Such
fluctuations are associated with the Galactic spiral density wave.
Earlier, we managed to estimate the spiral density wave parameters
quite reliably  from a lower number of masers
\citep{Bobylev2010,Bobylev2013}. Surprisingly, periodic
perturbations in $W$ velocities are visible in Fig.~\ref{f2} even
by eye.  In Fig.~\ref{f3}, power spectra of the radial $V_R$,
residual tangential $\Delta V_{circ}$ and vertical $W$ velocities
of masers are displayed. For application of the spectral analysis
described above, we adopt a four-armed spiral model
\citep{Bobylev2014}, $m=4$. The tangential and radial perturbation
amplitudes we obtained are $f_\theta=6.0\pm2.6$~km s$^{-1}$ and
$f_R=7.2\pm2.2$~km s$^{-1}$, respectively. The perturbation
wavelengths are $\lambda_\theta=3.2\pm0.5$~kpc and
$\lambda_R=3.0\pm0.6$~kpc. Note that all the error bars were found
using the Monte-Carlo statistical method. Significance levels of
all the spectrum peaks shown in Fig.~\ref{f3} are not lower than
0.995.

Based on our estimates of the wavelength $\lambda$, we found the
pitch angle for the four-armed spiral pattern ($m=4$) from
equation~(\ref{a-04}) for each component of the velocity
independently:
      $i_R=-13.4^\circ\pm2.7^\circ$,
 $i_\theta=-14.3^\circ\pm2.3^\circ$. These two values
are in a good agreement with the estimate $i=-13\pm1^\circ$ for
$m=4$, obtained earlier using spatial distribution of 73 masers
\citep{Bobylev2014}.

The most interesting result of this study is detection of a
periodic wave in vertical $W$ velocities versus distance $R$ which
is especially noticeable in the area of the Local arm and Perseus
arm~($R\approx9.5$~kpc). From our spectral analysis, we found the
following characteristics of this wave: the perturbation
wavelength is
 $\lambda=3.4\pm0.7$~kpc and the amplitude,
 $f_W=4.3\pm1.2$~km s$^{-1}.$

The waves found for all three velocity sets are shown in
Fig.~\ref{f2} (solid curves). The logarithmic nature of the waves
is clearly visible. Directly at the distance of the Sun, the top
of the radial velocity wave has a displacement in the direction to
the Galaxy center by $+19^\circ$; the shift of the tangential
velocities wave is $-11^\circ$ and that of the vertical velocities
$W$ wave is $+21^\circ.$ Usually we count the phase of the Sun
from the Carina--Sagittarius arm ($R\approx6.5$~kpc) according to
\cite{Rohlfs1977}. In this case, for the radial velocities we have
$(\chi_\odot)_R=-199^\circ\pm16^\circ$ and for the tangential
velocities, $(\chi_\odot)_\theta=-79^\circ\pm14^\circ.$

Note that the wave found in the tangential velocities contradicts
the simple theory~\citep{Lin1964}, as in the center of arms (for
example, the Carina--Sagittarius arm) these velocities should be
equal to zero. So we have a discrepancy with the theory amounting
to $\pi/2$. On the other hand, the obtained values of the phase of
the Sun are in a good agreement with the values found from data on
73 masers in \cite{Bobylev2013}, $-50^\circ\pm15^\circ$ and
$-160^\circ\pm15^\circ$ from the residual tangential and radial
velocities, respectively.

Now we present results of our solution of the Bottlinger's
equations which include perturbations from the spiral density wave
(\ref{DelVRot}) in accordance with the linear Lin\& Shu theory
\citep{Mish1979}, where amplitudes of radial and tangential
perturbations, pitch angle and phase  of the Sun are free
parameters. The results are the following: the amplitude of radial
perturbations is $f_R=6.0\pm1.4$~km s$^{-1}$; that of tangential
perturbations, $f_\theta=0.6\pm2.3$~km s$^{-1}$; the pitch angle
is $i=-11.7^\circ\pm4.9^\circ$ for the four-armed spiral model,
$m=4$; the phase of the Sun is $\chi_\odot=-118^\circ\pm22^\circ$.
We see that the amplitude of tangential perturbations is not
significant, although oscillations are clearly seen in the data
(Fig.~\ref{f2}). This result differs from that obtained using
spectral analysis, which can be explained by a discrepancy between
the data and Lin\& Shu model.

For the perturbations of vertical velocities $W$, there exists no
theory, but as it can be seen from Fig.~\ref{f2}, the wave in $W$
velocities is similar to the wave in radial velocities. Such a
result is in quite a good agreement with the results of numerical
simulations \citep{Faure2014,Debattista2014}.

\begin{figure}
 \includegraphics[width=1.0\columnwidth]{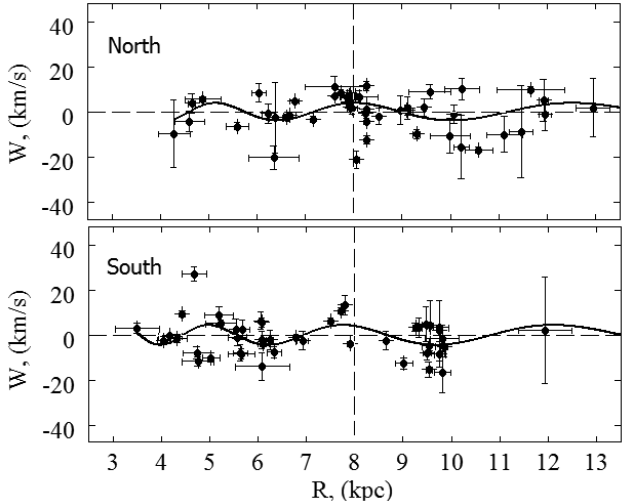}
 \caption{
The vertical $W$ velocities of  masers versus $R$.  The vertical
dashed line indicates the position of the Sun. }
  \label{f4}
 \end{figure}

As follows from the simulation results
\citep{Faure2014,Debattista2014}, the vertical velocities can be
inverted depending on the hemisphere. We checked this effect on
our sample of masers, though it is not so large. We divided the
whole sample in half --- into the North and South parts, and got
two solutions: $(f_W)_{North}=3.9\pm2.0$~km s$^{-1}$,
$\lambda_{North}=3.5\pm1.1$~kpc and $(f_W)_{South}=4.4\pm2.0$~km
s$^{-1}$, $\lambda_{South}=3.5\pm1.2$~kpc. Thus, we can see that
these solutions for amplitudes and wavelengths differ slightly
from those found using the whole data set. The phase shift between
the Northern and Southern waves is only about 20 degrees. We
conclude that the inversion is not detected. In Fig.~\ref{f4},
vertical velocities, together with the fitted waves (solid
curves), are shown for the galactic North and South hemispheres.

\section{CONCLUSIONS}\label{conclusions}
We have collected literature data on 119 Galactic masers with
known trigonometric parallaxes measured by means of VLBI. Based on
these series, we have re-determined the parameters of the Galactic
spiral density wave using the method of periodogram analysis.

We rejected 12 masers belonging to the Gould belt in order to
avoid the influence of its peculiar properties on the results of
the further analysis. The tangential and radial perturbation
amplitudes are $f_\theta=6.0\pm2.6$~km s$^{-1}$ and
$f_R=7.2\pm2.6$~km s$^{-1}$, respectively; the perturbation
wavelength is
 $\lambda_\theta=3.2\pm0.5$~kpc and
      $\lambda_R=3.0\pm0.6$~kpc.
Based on the derived values of $\lambda$, a new value of the
kinematic pitch angle for a four-armed spiral pattern is obtained:
$i=-13.9\pm2.5^\circ.$ The phase of the Sun $\chi_\odot$ in the
spiral density wave is $-79^\circ\pm14^\circ$ and
$-199^\circ\pm16^\circ$ from the residual tangential and radial
velocities, respectively (we measured the phase from the
Carina--Sagittarius arm).

The most interesting result of this work is detecting a wave in
the $W$ spatial velocities versus distance $R$ which is especially
noticeable in the area of the Local arm  and the Perseus arm. From
our spectral analysis, we have found the following characteristics
of this wave: the perturbation wavelength is
$\lambda_W=3.4\pm0.7$~kpc and the amplitude, $f_W=4.3\pm1.2$~km
s$^{-1}$. The wave obtained from $W$ velocities is closer to the
wave obtained from radial velocities. We found that similar wave
parameters can be derived both from the Northern and Southern
masers.

 \section*{ACKNOWLEDGEMENTS}
The authors are thankful to the anonymous referee for critical
remarks which permitted to improve the paper. This study was
supported by the ``Nonstationary Phenomena in Objects of the
Universe'' Program of the Presidium of Russian Academy of Sciences
(P--21). The authors are thankful to
 Nikolai Samus' for editing the text.

\end{document}